\documentclass[aps,prd,12pt,a4paper,groupedaddress,preprintnumbers,nofootinbib,floatfix]{revtex4-1}

\usepackage[top=2.9cm, bottom=2.1cm, left=2cm, right=2cm]{geometry}
\usepackage[english]{babel}
\usepackage{amsmath,amssymb,amsfonts, dsfont}
\usepackage{subfigure}
\usepackage{graphicx}
\usepackage{color}
\usepackage{hyperref}
\usepackage{dcolumn}
\usepackage{bm}
\usepackage{float}
\usepackage{mathtools}
\usepackage{mathrsfs}
\usepackage{amsthm}
\usepackage{bbm}
\usepackage{pxfonts}
\usepackage[vcentermath]{youngtab}
\usepackage[all]{xy}
\usepackage{accents}
\usepackage{cases}
\usepackage{comment}
\usepackage{placeins}
\usepackage{xspace}
\usepackage{cancel} 
\usepackage{slashed}
\usepackage{soul}
\usepackage{feynmp}
\usepackage{epstopdf}
\usepackage{xcolor}

\definecolor{orange}{cmyk}{0,0.5,1,0}
\definecolor{rossoCP3}{cmyk}{0,.88,.77,.40}
\definecolor{graa}{rgb}{0.8,0.8,0.8}
\definecolor{blaa}{rgb}{0.2,0.2,0.6}
\hypersetup{
	colorlinks,
	bookmarksopen,
	bookmarksnumbered,
	citecolor=blaa, 		
	linkcolor=rossoCP3,	
	urlcolor=rossoCP3,			
	    }

\newcommand{\be}{\begin{equation}}
\newcommand{\ee}{\end{equation}}

\newcommand{\SU}{\mathrm{SU}}

\newcommand{\SO}{\mathrm{SO}}
\newcommand{\Tr}{\text{Tr}}

\newcommand{\TeV}{\text{TeV}}

\begin{document}


\title{\texorpdfstring{\Large\color{rossoCP3} The high energy fate of the minimal Goldstone Higgs}{}}
\author{Helene {\sc Gertov}}
\email{gertov@cp3.sdu.dk}
\author{Sofie {\sc Gregersen}}
\email{gregersen@cp3.sdu.dk}
\author{Francesco {\sc Sannino}}
\email{sannino@cp3.dias.sdu.dk}
\affiliation{{\color{rossoCP3} {CP}$^{ \bf 3}${-Origins}} \& the Danish Institute for Advanced Study {\color{rossoCP3}\rm{Danish IAS}},  University of Southern Denmark, Campusvej 55, DK-5230 Odense M, Denmark.}
\author{Kimmo {\sc Tuominen}}
\email{kimmo.i.tuominen@helsinki.fi}
\affiliation{Department of Physics, University of Helsinki, \& Helsinki Institute of Physics, P.O.Box 64, FI-00014 University of Helsinki, Finland}

\begin{abstract}
We consider a minimal model where the Higgs boson arises as an
elementary pseudo-Nambu-Goldstone boson. The model is based on
an extended scalar sector with global SO(5)/SO(4) symmetry. To
achieve the correct electroweak symmetry breaking pattern, the model is
augmented either with an explicit symmetry breaking term or an extra
singlet scalar field. We consider separately both of these possibilities.
We fit the model with the known
particle spectrum at the electroweak scale and extrapolate to high
energies using renormalization group. We find that the model can
remain stable and perturbative up to the Planck scale provided that
the heavy beyond Standard Model scalar states have masses
in a narrow interval around 3 TeV.
\\[.3cm]
{\footnotesize  \it Preprint: CP$^3$-Origins-2017-050 DNRF90 \& HIP-2018/2-TH} 
\end{abstract}
\maketitle
\newpage

\section{Introduction}

The discovery of the Higgs boson at the LHC has verified the Standard Model
(SM)-like pattern of electroweak symmetry breaking. A possible interpretation
of this discovery is to take the SM particle content (possibly
extended by three right handed neutrinos~\cite{Asaka:2005an,Asaka:2005pn}) to describe all elementary particle
interactions below the Planck energy: with the observed Higgs mass the scalar
self coupling does not develop a Landau pole and while the scalar self
coupling runs negative around scale $10^{10}$ GeV, this results only in a
metastability of the low energy vacuum~\cite{Degrassi:2012ry,Antipin:2013sga}.

Even so, if the inflationary scale is high enough, one must explain why the Higgs
field settled into the false low energy vacuum in spite of large field
excursions induced by the inflationary fluctuations~\cite{EliasMiro:2011aa,Kobakhidze:2013tn,Fairbairn:2014zia}.
In extensions of the SM with larger scalar sectors this problem can be
alleviated~\cite{Gabrielli:2013hma,Bhattacharya:2014gva},
as the presence of additional bosonic degrees of freedom
coupling only with the Higgs can overcome the SM contribution of the top quark.
However, with a
larger scalar sector involving more couplings another problem emerges as one or
more of these couplings can develop Landau poles below the Planck scale.

These basic features, following from the renormalization group evolution of
the scalar self couplings, are very sensitive to
the degrees of freedom and the relative strengths of their couplings within
the scalar sector. Therefore, in the absence of direct signal of any
new resonance, the vacuum stability and perturbativity of the couplings provide
essential theoretical constraints for various BSM scenarios.

An interesting class of SM extensions, where compatibility of the spectrum with
the one observed at LHC can be achieved, is the one where the Higgs boson
arises as a pseudo-Nambu-Goldstone Boson (pNGB)~\cite{Schmaltz:2005ky}.
In this type of models, the Higgs sector can be either elementary or composite.
The composite case has been much studied as it
allows to address the hierarchy problem. However, it lacks simple dynamics
to produce the SM-fermion masses. The elementary case, on
the other hand, provides a calculable framework
to assess the observed symmetry-breaking pattern
and low-energy spectrum ~\cite{Coleman:1973jx}, and facilitates an effective
description of flavour physics in terms of
Yukawa interactions as in the SM~\cite{Alanne:2016mmn}. Moreover,
the vacuum expectation value of the field in the elementary
realization does not provide a cutoff for the model. Instead, the model
with elementary scalars can be taken as an effective description valid all
up to the Planck energy. We will therefore follow
the latter route in this paper, and investigate the running of the
couplings and the fate of the vacuum in elementary Goldstone Higgs models.

%
%

In this paper we consider explicitly
the minimal symmetry breaking patterm, $\SO(5)\to \SO(4)$, where the Higgs
emerges as a pNGB.
It has been showed \cite{Alanne:2016mmn}
that the minimal particle content of $\SO(5)\to \SO(4)$ needs to be extended
in order to have a non-trivial vacuum.
There are two basic extensions: One is by adding an explicit breaking term which
couples to the singlet and the other is by adding a new scalar which couples to
the $\SO(5)\to \SO(4)$ scalar multiplet. We will separately consider
these two possibilities.

The paper is organised as follows: In section \ref{sec:model} we
review briefly the $\SO(5)\to\SO(4)$ model where the Higgs is an elementary
pseudo Goldstone Boson and determine the $\beta$-functions.
In section \ref{sec:exp_break} we  examine  the running of the couplings in
the case with an explicit breaking term. In section \ref{sec:scalar} we examine
the running of the couplings in the case with an extra  scalar and
in section \ref{sec:conclusion} we present our conclusions.

\section{The Minimal Model}
\label{sec:model}
The minimal extension of SM leading to pNGB Higgs can be written
as a linear $\sigma$-model over the coset SO(5)/SO(4).
The general SO(5) invariant potential in terms of SO(5) vector $\Sigma$ is
\begin{equation}
	V_0=\frac{m^2}{2}\Sigma^\dagger\Sigma+\frac{\lambda}{4!}(\Sigma^\dagger\Sigma )^2.
	\label{eq:Vcom}
\end{equation}
The electroweak gauge group is identified within the SU(2)$_L\times$SU(2)$_R$
subgroup of SO(4). Then the vacuum of the theory can be parametrized as a
superposition between a vacuum which preserves the electroweak symmetry, $E_0=(0,0,0,1,0)^T$, and
a vacuum which breaks the electroweak symmetry, $E_B=(0,1,0,0,0)^T$, as
\begin{equation}
	E_\theta=\cos\theta\, E_0+\sin\theta\, E_B,
\end{equation}
The $\SO(5)$ scalar multiplet can then be written as
\begin{equation}
	\Sigma=\left( \sigma+i \pi_a X_\theta^a\right)E_\theta,
\end{equation}
where $X_\theta^a$ are the broken generators of the $\SO(5)\to \SO(4)$ and can
be found in appendix \ref{app:X5}, $\pi_a$ are the Goldtone Bosons and
$\sigma$ is a masive scalar field and the only field which obtains a nonzero vacuum expectation value (vev).

\begin{figure}
	\centering
	\includegraphics[scale=0.7]{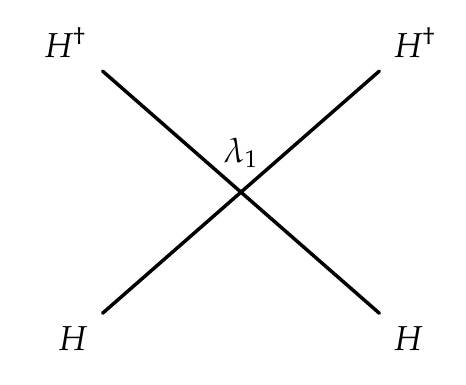}\quad
	\includegraphics[scale=0.7]{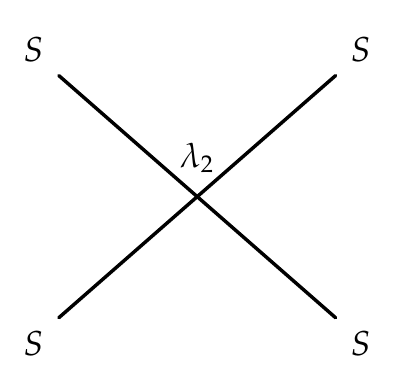}\quad
	\includegraphics[scale=0.7]{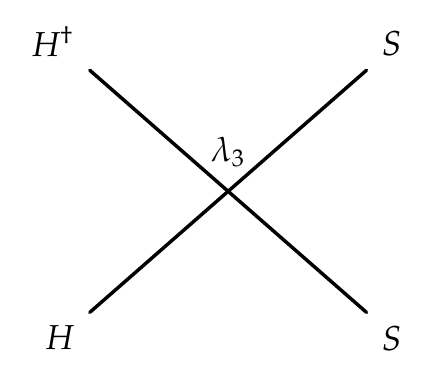}
	\caption{shows the three different three-level Feynman diagrams for the $\SO(5)\to \SO(4)$ theory.}
	\label{fig:Feynman_so5}
\end{figure}

The scalar multiplet can be parametrized in many ways.
For our purposes, it is most convenient to rewrite it in the basis of
eigenstates under the electroweak interactions, i.e. a complex
doublet with a neutral and a charged component
and a real scalar singlet.
In terms of the $\sigma$ field and the Goldstone Bosons,
the  doublet and the singlet are:
\begin{equation}
	H=\frac{1}{\sqrt{2}}\begin{pmatrix}
		\pi_1+ i \pi_2\\
		\sigma\sin\theta+\pi_4\cos\theta+i \pi_3
	\end{pmatrix}
	\qquad\text{and}\qquad
	S=\sigma\cos\theta-\pi_4\sin\theta.
\end{equation}
In this basis the higher order potential is\footnote{Note that at tree level the parameters are matched with $\lambda$ and $m$:
$\lambda_1 = \frac{\lambda}{3!}$, $\lambda_2 = \frac{\lambda}{4!}$,
$\lambda_1 = \frac{\lambda}{3!}$, $m_h=m$ and $m_s=\frac{m}{\sqrt{2}}$,
and the tree level potential matches with the one given in \eqref{eq:Vcom}.}
\begin{equation}
	V=m_h^2H^\dagger H+m_s^2 S^2+\lambda_1 \left(H^\dagger H\right)^2+\lambda_2 S^4+\lambda_3 H^\dagger H S^2.
	\label{eq:VHS}
\end{equation}
The three couplings introduced in Eq.~\eqref{eq:VHS}
and represented in figure~\ref{fig:Feynman_so5} are derived, at tree level,
from the single coupling $\lambda$ appearing in Eq.~\eqref{eq:Vcom}. However,
they will run differently due to different higher order contributions to each
coupling $\lambda_1$, $\lambda_2$ and $\lambda_3$.

The stability constraints on the scalar couplings are
\begin{equation}
	\lambda_1\geq 0,\qquad \lambda_2\geq0,\qquad \lambda_3+2\sqrt{\lambda_1\lambda_2}\geq0.
	\label{eq:stability}
\end{equation}
where $\lambda_1$ and $\lambda_2$ are strictly positive, but $\lambda_3$ is allowed
to take negative values within the bound implied by the above equation.


The gauge interactions are determined, as in~\cite{Alanne:2014kea,Gertov:2015xma}, from the kinetic term:
\begin{equation}
	\mathcal{L}_\text{kin}=\left(D_\mu \Sigma\right)^\dagger D^\mu \Sigma ,
\end{equation}
where the covariant derivative is
\begin{equation}
	D_\mu=\partial_\mu \Sigma -i g_w W^aT^a_L\Sigma+i g_y B T^3_R\Sigma.
\end{equation}
The $T_L$ and $T_R$  are the generators of the $\SU_L(2)$ and $\SU_R(2)$ respectively and
they are explicitly defined in Appendix~\ref{app:X5}.

The Lagrangian for the Yukawa couplings is
\begin{eqnarray}
	\mathcal{L}_\text{Yuk}^\text{top}= y_t (Qt^c)^\dagger_\alpha P_\alpha \Sigma ,
\end{eqnarray}
where $\alpha$ is a $\SU_L(2)$ index and $P_\alpha$ are pseudo-projectors defined as
\begin{equation}
	P_1=\frac{1}{\sqrt{2}}\big( 0,\,\,0,\,\,1,\,\,i,\,\,0 \big)\qquad\text{and}\qquad P_2=\frac{1}{\sqrt{2}}\big( 1,\,\,i,\,\,0,\,\,0,\,\,0 \big) \, .
\end{equation}
These pseudo-projectors pick the correct parts of the doublets in $\Sigma$.
We will consider here only the Yukawa interactions of the top quark as these
are the ones giving the dominant contribution from fermions
to the beta functions.

The beta functions describing the running of the couplings above the
renormalization scale $\mu_0$ are computed using the standard
methods~\cite{Machacek:1983fi,Machacek:1983tz,Machacek:1984zw}.
The beta functions at one-loop order for the three scalar couplings in this theory are
\begin{equation}
\begin{split}
	\beta_{\lambda_1}&=\frac{1}{(4\pi)^2}\left( \frac{9 }{8}g_w^4+\frac{3 }{4}g_w^2 g_y^2 +\frac{3 }{8}g_y^4 -9 g_w^2 \lambda_1-3 g_y^2\lambda_1-6y_t^4+12\lambda_1 y_t^2+24 \lambda_1^2+2 \lambda_3^2\right)\\
	\beta_{\lambda_2}&=\frac{1}{(4\pi)^2}\left( 72 \lambda_2^2+2 \lambda_3^2\right)\\
	\beta_{\lambda_3}&=\frac{1}{(4\pi)^2}\left(-\frac{9}{2}  g_w^2 \lambda_3-\frac{3}{2}g_y^2 \lambda_3+6y_t^2\lambda_3 +12 \lambda_1 \lambda_3+24\lambda_2 \lambda_3+8 \lambda_3^2\right).
\end{split}
\end{equation}
However, there is an important caveat that we need to take into account now.
In \cite{Alanne:2016mmn} some of the authors of the present paper found that the gauge and Yukawa
interactions are not enough to align the vacuum away from zero in any
$\SO(N)\to\SO(N-1)$ theory where the Higgs is an elementary pNGB.  Two ways of solving this issue were put forward in \cite{Alanne:2016mmn}: First, by adding a small
explicit breaking term competing with the one loop potential contribution, and second, by adding an extra scalar field
which couples to $\Sigma$ via a portal coupling. Within the second approach three
new couplings need to be introduced. 

In the following two section we will answer the relevant question of what is the behaviour of the running of
all the couplings in both scenarios at high energy. To our knowledge these is the first comprehensive analysis  of these theories at short distances.

\section{$\SO(5) \to \SO(4)$ + explicit breaking term}\label{sec:exp_break}

In this section we consider the tree level potential given in eq. (\ref{eq:Vcom}). At some renormalization scale, $\mu_0$, the three scalar couplings are assumed to combine so that the potential is SO(5) invariant.
Furthermore, assuming perturbative values of the couplings, the one-loop  corrections are computable using the Coleman-Weinberg potential which is defined as
\begin{equation}
\delta V(\Phi) = \frac{1}{64 \pi^2} \text{Str} \left[ \mathcal{M}_0^4(\Phi) \left(\log \frac{\mathcal{M}_0^2(\Phi)}{\mu_0^2}-C \right) \right] + V_{GB} \, ,\label{eq:CW}
\end{equation}
where $\mathcal{M}_0$ is the mass matrix, $\text{Str}$ is the supertrace
where the sums over scalar, fermion and vector degrees of freedom are weighted
with factors $1$, $-2$ and $3$, respectivley. The constant $C$ depends on the
particle type and is $C=3/2$ for scalars and fermions and $C=5/6$ for gauge
bosons. The factor $V_{GB}$ represents the one-loop corrections from
the Goldstone bosons which we will neglect since their contributions to the
potential are much smaller than the correction from massive particles.

As argued above we can ensure existence of a nontrivial vacuum by adding an explicit symmetry breaking term, given by
\begin{equation}
V_B = C_B v^3 S \, .
\end{equation}
On the basis of the results in~\cite{Alanne:2016mmn} we take $C_B> 0$ and
small, in order to have a small breaking. Concretely, we consider
$0 < C_B < 0.1$. Imposing the correct mass of the Higgs and  minimizing of the potential wrt. $\theta$ give us an upper and lower bound on the mass $M_\sigma$
which is $1.35 \, \TeV < M_\sigma <3.20 \, \TeV$. This corresponds to a
coupling $\lambda$ in the interval $0.29 < \lambda < 1.68$. These are the
values below the renormalization scale $\mu_0$, where we can describe the
theory through the Coleman-Weinberg potential. The renormalization scale
$\mu_0$ is a function of $M_\sigma$ and in the allowed interval of $M_\sigma$
it is nearly constant with the value $\mu_0 \approx 2 \, \TeV$.

\begin{figure}
	\includegraphics[scale=0.5]{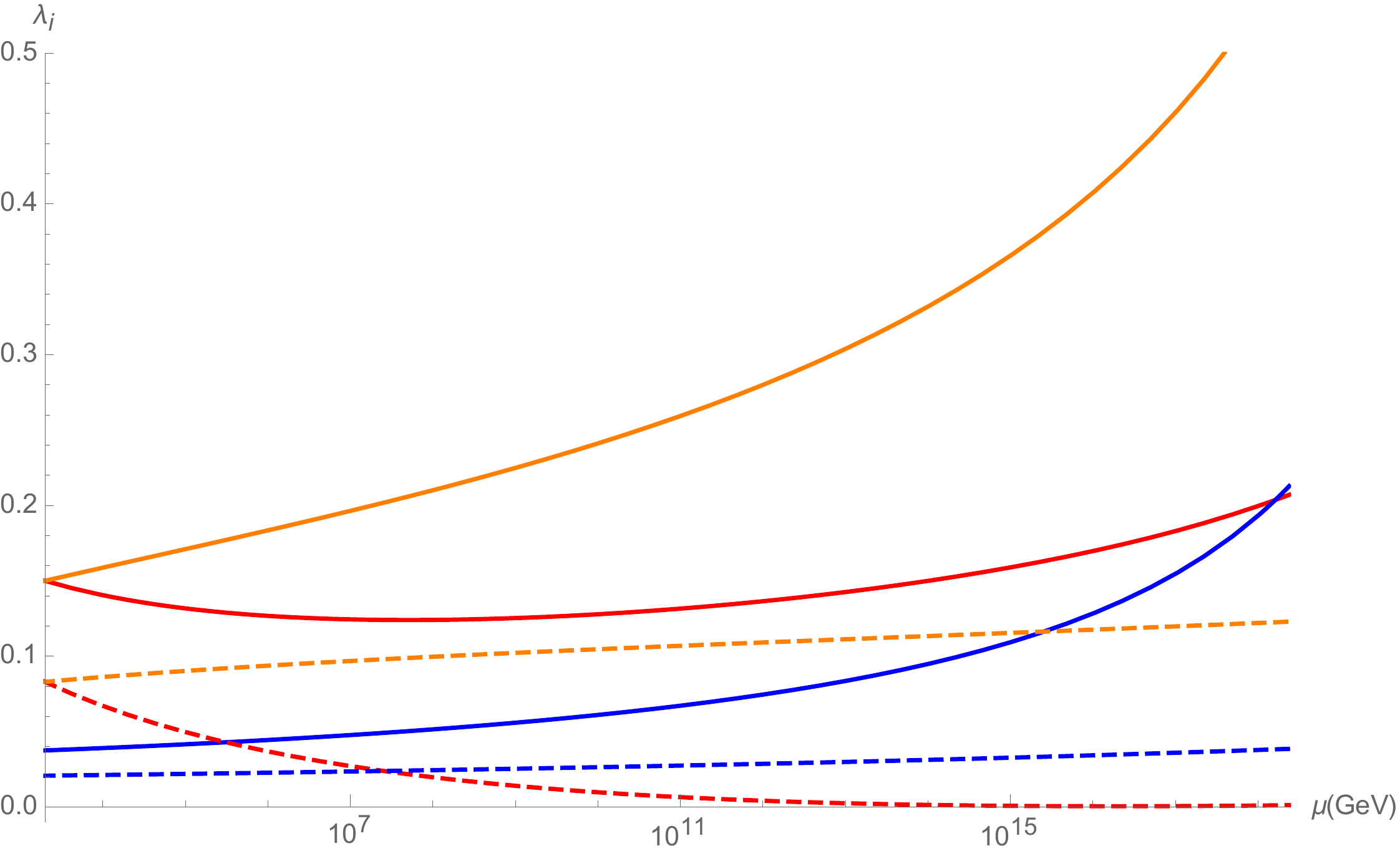}
	\caption{The figure shows the running of the couplings above 2 TeV, where the dashed lines correspond to the lower bound on the combined $\lambda$ and the solid lines correspond to the upper bound. The red lines are $\lambda_1$, the blue lines are $\lambda_2$ and the orange lines are $\lambda_3$. The initial values of the three running couplings all depend on the same coupling $\lambda$. On the plot $\lambda_2$ does however start in a value different from the two other couplings, this is because of the different normalizations.}
	\label{fig:CBlambdamaxmin}
\end{figure}

Above the renormalization scale the coupling $\lambda$ splits into three
couplings which run according to their respective beta functions. At the
renormalization scale the value of all three couplings is determined by
$\lambda$. The couplings $\lambda_1$ and $\lambda_2$ must be positive and
we require that all couplings remain free of Landau poles
all the way to the Planck scale. From these constraints we
find $0.50 < \lambda < 0.90$ which corresponds to a sigma mass in the interval
$2.93 \, \TeV < M_\sigma < 3.10 \, \TeV$  and $\sin \theta$ in the interval
$0.034 < \sin \theta < 0.043$. Finally the vacuum is found to be between
$5.66 \, \TeV < v < 7.18 \, \TeV$.
These constraints are illustrated in figure~\ref{fig:CBlambdamaxmin}. The figure show the scale evolution of the couplings $\lambda_1$ (red),
$\lambda_2$ (orange) and $\lambda_3$ (blue), and the
constraints restrict the running of the couplings to
lie between the corresponding solid and dashed curves.

\begin{figure}
	\centering
	\includegraphics[scale=0.45]{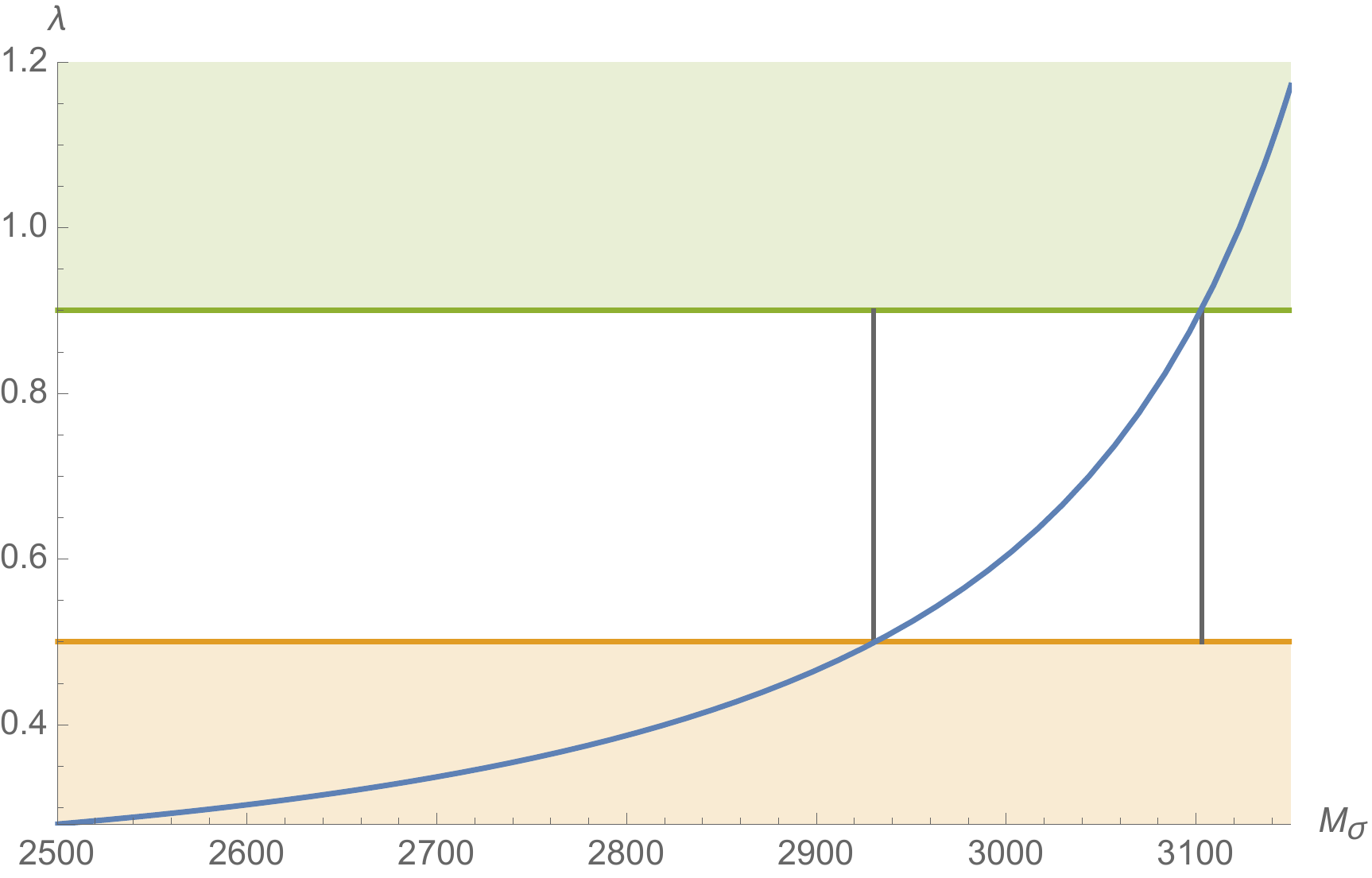}
	\includegraphics[scale=0.54]{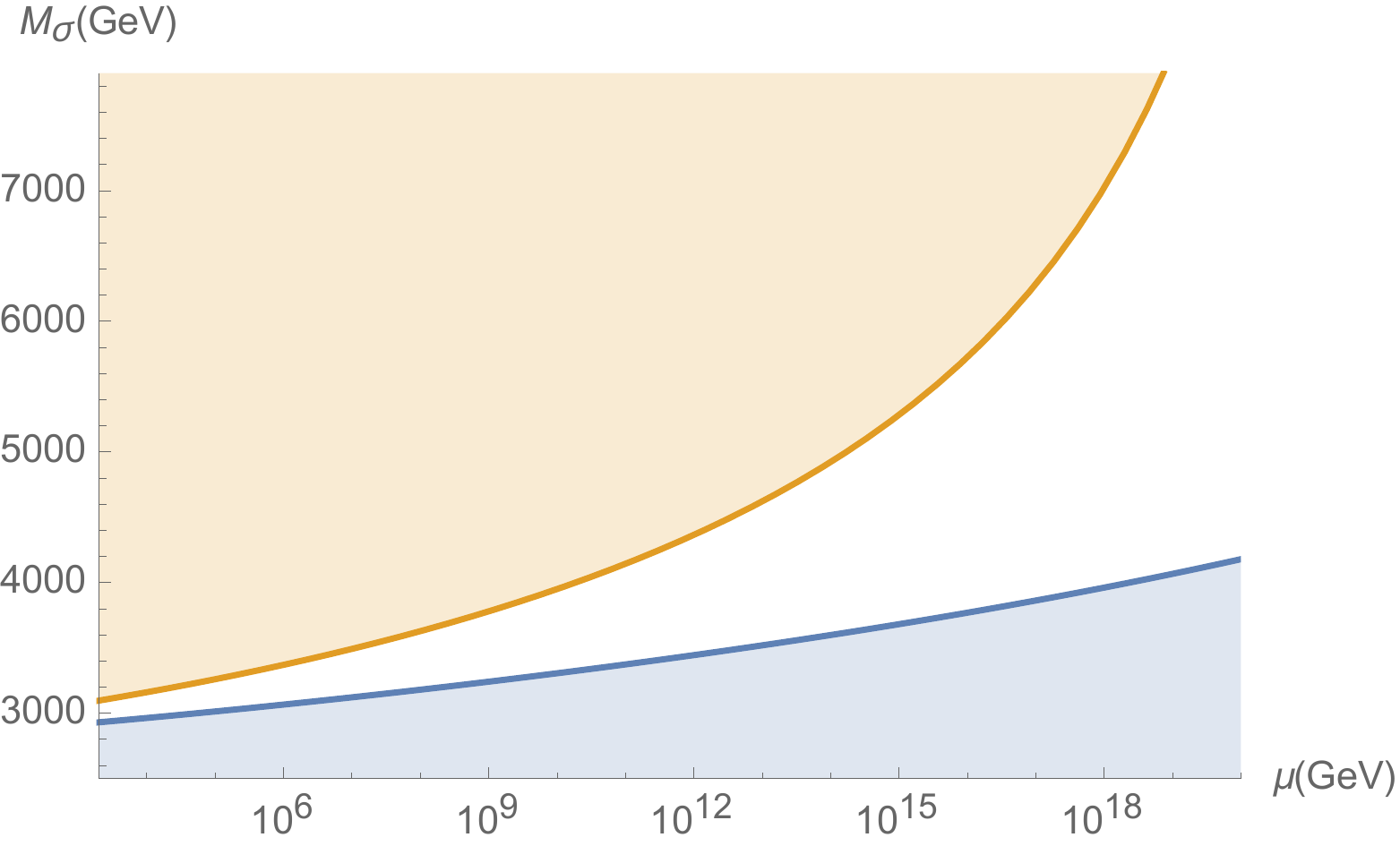}
	\caption{\textbf{Left panel:} The blue line shows the relationship between $\lambda$ and $M_\sigma$ below $\mu_0 = 2 \, \TeV$ when the correct Higgs mass is imposed. The green area is excluded due to Landau poles and the orange area is excluded due to stability. This gives a narrow window for $M_\sigma \in [2.93,3.10] \, \TeV$ indicated by the grey vertical lines. \textbf{Right panel:} Running of $M_{\sigma}$. The shaded blue area is excluded due to stability and the orange area due to Landau poles. The true running of $M_\sigma$ lies on a curve in the white region and depends on the value of $\lambda$ at $\mu_0=1.70 \, \TeV$.}
	\label{fig:CBstability}
\end{figure}

The relationship between $\lambda$ and $M_\sigma$ below $\mu_0 = 2 \, \TeV$ (i.e. where the coupling does not run) is plotted in the left panel of figure \ref{fig:CBstability}.
The blue line shows the relation between $M_\sigma$ and $\lambda$ when the
potential is minimised with respect to $\sin\theta$ and the correct mass of the
Higgs is imposed.
The shaded regions are excluded due to perturbativity  and stability
constraints on $\lambda$ and the grey vertical lines correspond to
$M_\sigma=2.93 \, \TeV$ and $M_\sigma=3.10 \, \TeV$ respectively.
The right panel of figure \ref{fig:CBstability} shows the mass of the
$\sigma$ particle, $M_\sigma$, as function of the renormalization scale $\mu$.
The blue shaded region is excluded due to vacuum stability as the
couplings run.
The orange shaded region is excluded due to the perturbativity  of the couplings (i.e. the absence of Landau poles below the Planck scale).
This means that the running of $M_\sigma$ lies on a curve in the
white region\footnote{Note: when generating this plot, we have assumed that $\sin \theta$ is constant for all $\mu$. We can make this assumption since $\sin \theta$ is generated from a one-loop diagram and therefore the corrections to $\sin \theta$ will be of second order.}.

As in the Elementary Goldstone Higgs model, first proposed in
\cite{Alanne:2014kea,Gertov:2015xma}, the observed Higgs boson is a
superposition between the $\sigma$ and the $\pi_4$ particles. This
superposition can be described by a mixing angle $\alpha$.
For small values of $\theta$ this can be approximated as
\begin{equation}
	\begin{split}
		\alpha\approx &\frac{\pi}{2}
	-\frac{6 \theta^3 v^4 }{54 M_\sigma^8+192 \pi ^2 M_\sigma^6 v^2-11 C_B M_\sigma^4 v^4}\\
	&\qquad\quad \left[6 B M_\sigma^4-A \left(C_B v^4-13 M_\sigma^4+6 M_\sigma^4 \log \frac{M_\sigma^2}{\theta ^2 v^2}\right)\right] +\mathcal{ O} (\theta^5 ) ,
	\end{split}
\end{equation}
where A and B are coefficients depending only on gauge and Yukawa couplings. They are given by
\begin{equation}
	\begin{split}
		A=& 3 \left(\frac{g_w^4}{8}+\frac{1}{16} \left(g_w^2+g_y^2\right)^2\right)-3 y_t^4 ,\\
		B=& \frac{3}{16} \left[\left(g_w^2+g_y^2\right)^2 \left( \log \frac{g_w^2+g_y^2}{4} -\frac{5}{6}\right)+2 g_w^4 \left( \log \frac{g_w^2}{4}-\frac{5}{6}\right)\right]-3 y_t^4 \left(2 \log \frac{y_t^2}{2} -3\right).
	\end{split}\label{eq:AB}
\end{equation}
If $\alpha$ is close to $\frac{\pi}{2}$ the observed Higgs is mostly the
Goldstone Boson, $\pi_4$, while the observed Higgs is mostly the scalar
$\sigma$ if $\alpha$ is close to zero. In the interval for $M_\sigma$ found
above, $\alpha$ is very close to $\frac{\pi}{2}$ and the observed Higgs is
almost the Goldstone boson.
Defining the two mass eigenstates  $h_1=\cos\alpha \, \sigma+\sin\alpha \, \pi_4$ and $h_2=-\sin\alpha \, \sigma+\cos\alpha \, \pi_4$,
where $h_1$ is the observed Higgs, we can
calculate the self couplings of the physical mass eigenstates:
\begin{equation}
	\begin{split}
		\lambda_{h_1h_1h_1}= \frac{3 M_\sigma^2 \cos \alpha}{v},\qquad
		\lambda_{h_1h_1h_2}=  \frac{ M_\sigma^2 \sin \alpha}{v},\qquad
		\lambda_{h_1h_2h_2}= \frac{ M_\sigma^2 \cos \alpha}{v},\\
		\lambda_{h_2h_2h_2}= \frac{3 M_\sigma^2 \sin \alpha}{v},\qquad
		\lambda_{h_1h_1h_1h_1}= \frac{3 M_\sigma^2 }{v^2},\qquad
		\lambda_{h_1h_1h_1h_2}=0,\\
		\lambda_{h_1h_1h_2h_2}= \frac{ M_\sigma^2 }{v^2},\qquad
		\lambda_{h_1h_2h_2h_2}=0,\qquad
		\lambda_{h_2h_2h_2h_2}= \frac{3 M_\sigma^2 }{v^2}.
	\end{split}\label{eq:sefcoupling}
\end{equation}
Note that the quartic couplings do not depend on the mixing angle $\alpha$ and
the quartic couplings for $h_1$ and $h_2$ are identical.
However, the trilinear couplings do depend on $\alpha$. When $\alpha$ is close
to $\frac{\pi}{2}$ the trilinear self coupling of $h_1$ is very small while the
trilinear self coupling of $h_2$ is large. In the interval on the mass
$M_\sigma$ found above, we can calculate the ratio of the two trilinear couplings with the trilinear coupling of the SM:
\begin{eqnarray}
	1.02 \cdot 10^{-3}< \frac{\lambda_{h_1h_1h_1}}{\lambda_{hhh}^{SM}}<2.33\cdot 10^{-3} \qquad\text{and}\qquad
	23.5 > \frac{\lambda_{h_2h_2h_2}}{\lambda_{hhh}^{SM}}>20.9.
\end{eqnarray}
The trilinear coupling for $\lambda_{h_1h_1h_1}$ is very small compared to the corresponding coupling in SM and it is smallest when the $M_\sigma$ is smallest. However the behavior of $\lambda_{h_2h_2h_2}$ is the opposite: $\lambda_{h_2h_2h_2}$ is larger than the SM one and it is largest when $M_\sigma$ is smallest. 
 
We have therefore shown that this scenario leads to a viable elementary Goldstone Higgs framework valid up to the Planck scale. 

\section{$\SO(5)\to \SO(4)$ + an extra scalar $\Omega$}\label{sec:scalar}
Now we turn to the other possibility, namely adding an extra scalar $\Omega$ which couples to $\Sigma$ via a portal coupling. We assume that $\Omega$ is both real and has a $Z_2$ symmetry. The new tree level potential is then
\begin{equation}
	V_0=\frac{m^2}{2} \Sigma^\dagger\Sigma+\frac{m_\Omega^2}{2} \Omega^2+\frac{\lambda}{4!}\left(\Sigma^\dagger\Sigma\right)^2 +\frac{\tilde\lambda}{4} \Sigma^\dagger\Sigma\Omega^2 +\frac{\tilde\lambda_\Omega}{4!}\Omega^4 .
	\label{eq:fullpotential1}
\end{equation}
As in the previous section, we write the potential in terms of the
doublet and singlet fields:
\begin{equation}
	V=m_h^2H^\dagger H+m_s^2 S^2+m_\Omega^2 \Omega^2 +\lambda_1 \left(H^\dagger H\right)^2+\lambda_2 S^4+\lambda_3 H^\dagger H S^2 +\lambda_4 H^\dagger H \Omega^2+\lambda_5 S^2\Omega^2+\lambda_\Omega \Omega^4 \, .
\label{eq:fullpotential2}
\end{equation}
Here the couplings $\lambda_1$, $\lambda_2$ and $\lambda_3$ are, at tree level,
determined by the self-interaction $\lambda$ in the SO(5) invariant potential,
Eq.~\eqref{eq:fullpotential1}, and similarly $\lambda_4$ and $\lambda_5$ are determined by $\tilde{\lambda}$. However, similarly to the situtation treated
in the previous section, all these couplings will receive different contributions at one-loop order and, consequently, their running will be different.
The beta functions for the six different couplings are:
\begin{equation}
\begin{split}
	\beta_{\lambda_1}&=\frac{1}{(4\pi)^2}\left( \frac{9 }{8}g_w^4+\frac{3 }{4}g_w^2 g_y^2 +\frac{3 }{8}g_y^4 -9 g_w^2 \lambda_1-3 g_y^2\lambda_1-6y_t^4+12\lambda_1 y_t^2+24 \lambda_1^2+2 \lambda_3^2+2\lambda_4^2 \right) , \\
	\beta_{\lambda_2}&=\frac{1}{(4\pi)^2}\left( 72 \lambda_2^2+2 \lambda_3^2+2\lambda_5^2 \right) ,\\
	\beta_{\lambda_3}&=\frac{1}{(4\pi)^2}\left(-\frac{9}{2}  g_w^2 \lambda_3-\frac{3}{2}g_y^2 \lambda_3+6y_t^2\lambda_3 +12 \lambda_1 \lambda_3+24\lambda_2 \lambda_3+8\lambda_3^2+4\lambda_4\lambda_5\right) ,\\
	\beta_{\lambda_4}&=\frac{1}{(4\pi)^2}\left(-\frac{9}{2}g_w^2\lambda_4-\frac{3}{2}g_y^2\lambda_4+6y_t^2\lambda_4+12\lambda_1\lambda_4+8\lambda_4^2+4\lambda_3\lambda_5+24\lambda_4\lambda_\Omega\right) ,\\
	\beta_{\lambda_5}&=\frac{1}{(4\pi)^2}\left(4\lambda_3\lambda_4+24\lambda_2\lambda_5+16\lambda_5^2+24\lambda_5\lambda_\Omega\right) ,\\
	\beta_{\lambda_\Omega}&=\frac{1}{(4\pi)^2}\left(2\lambda_4^2+2\lambda_5^2+72\lambda_\Omega^2\right) .
\end{split}
\end{equation}
Note the similarity between the beta functions of $\lambda_3$ and $\lambda_4$, which are the couplings between the scalar doublet and the two scalar singlets.
Also note the similarity between the beta functions  of $\lambda_2$ and $\lambda_\Omega$, which are the self-couplings of the two singlets:
The new scalar singlet $\Omega$ has interactions which are analogous to the original singlet component $S$ of the SO(5) mutiplet $\Sigma$.

For simplicity we consider the case, where the mass of the new particle $\Omega$ is the same as the mass of $\sigma$.
Just as in the previous case, below the renormalization scale $\mu_0$ the couplings combine and the $\SO(5)\to \SO(4)$ symmetry is intact. Since the couplings are perturbative, we can use the Coleman-Weinberg potential introduced in~\eqref{eq:CW}. However, in this case the mixing  $\alpha$ is simpler and given by
\begin{eqnarray}
	\alpha\approx \frac{\pi}{2}-\frac{v^4\theta^3}{9M_\sigma^4+32\pi M_\sigma^2 v^2+\tilde\lambda v^4}\left[ 6B-A\left( 6 \log\frac{M_\sigma^2}{v^2\theta^2}-13 \right)\right]+\mathcal{ O} (\theta^5 )
\end{eqnarray}
where $A$ and $B$ are given in \eqref{eq:AB}.

As in the case treated in the previous section, we will now constrain the model
by requiring that correct spectrum of physical states is reproduced at low energies, and that the vacuum remains stable at high energies.
We also require that the one loop running of the dimensionless couplings is
free of Landau poles below the Planck scale. We treat separately the constraints relevant below and above the renormalization scale.

First, below the renormalization scale, we impose the correct mass of the Higgs and demand that the potential is minimized with respect to $\theta$. We find
that the common mass of
$\sigma$ and $\Omega$ has a minimum value: $M_\sigma=M_\Omega=2.80$ TeV.
Requiring that $\lambda$ remains perturbative, $\lambda\le 4\pi$, we find that the maximal value of the mass is 3.03 TeV. The minimal value of the mass corresponds to minimal values for the couplings $\lambda=0$ and $\tilde\lambda=0.01$ at the renormalization scale $\mu_0$.
At this value of $\lambda$ the renormalization scale is $\mu_0= 1.70$ TeV.

Second, above the renormalization scale the couplings run and, consequently, the
allowed mass ranges will be refined by constraints
due to stability of the potential and absence of Landau poles below the Planck
scale.

For the potential to be stable, the coupling $\lambda_1$ has to be positive all the way to the Planck scale. This requires that the combined coupling evaluated at the renormalization scale is $\lambda\geq 0.51$. The corresponding value for the renormalization scale is $\mu_0=1.70$ TeV. The lower limit of $\lambda$ corresponds to a mass of 2.81 TeV. Similarly the value of $\tilde\lambda$
at the renormalization scale is $\tilde\lambda=0.11$, while $\tilde\lambda_\Omega$ is a free variable with the only requirement that it cannot be negative.

Requiring that all the couplings remain perturbative all the way to the
Planck scale,
i.e. that there are no Landau poles, constrains the values of
the couplings evaluated at the renormalization scale from above.
We find a maximal value $\lambda=0.96$, which corresponds to a mass of 2.82
TeV. At this value of the mass, $\tilde\lambda=0.20$ and does not acquire a
Landau pole at energies below the Planck scale.
The renormalization scale in this case is 1.71 TeV. Again
$\tilde\lambda_\Omega$ is a free variable and has a maximum value of 0.90 at the
renormalization scale, when we require that it does not develop a Landau pole below the Planck scale.

The above constraints are summarized as follows:
when we analyse the running of the couplings we find an interval for the couplings $\lambda\in \lbrack 0.51,0.96\rbrack$, $\tilde\lambda\in\lbrack 0.11,0.20\rbrack$ and $\tilde\lambda_\Omega\in\lbrack 0,0.90\rbrack$. These intervals are quite narrow and hence restrict the scalar mass to a very narrow range: $M_\sigma=M_\Omega\in\lbrack 2.81,2.82\rbrack$ TeV. This scalar mass corresponds to $\sin\theta\in\lbrack 0.036,0.049\rbrack$ and $v\in\lbrack 4.98 , 6.76 \rbrack$ TeV. Finally, the renormalization scale  in this case must be in the interval $\mu_0\in\lbrack 1.70,1.71\rbrack$ TeV. Hence, we find that  $\mu_0\approx 1.7$ TeV, which is of similar magnitude as in the case of an explicit breaking term treated in the previous section.

\begin{figure}
	\includegraphics[scale=0.6]{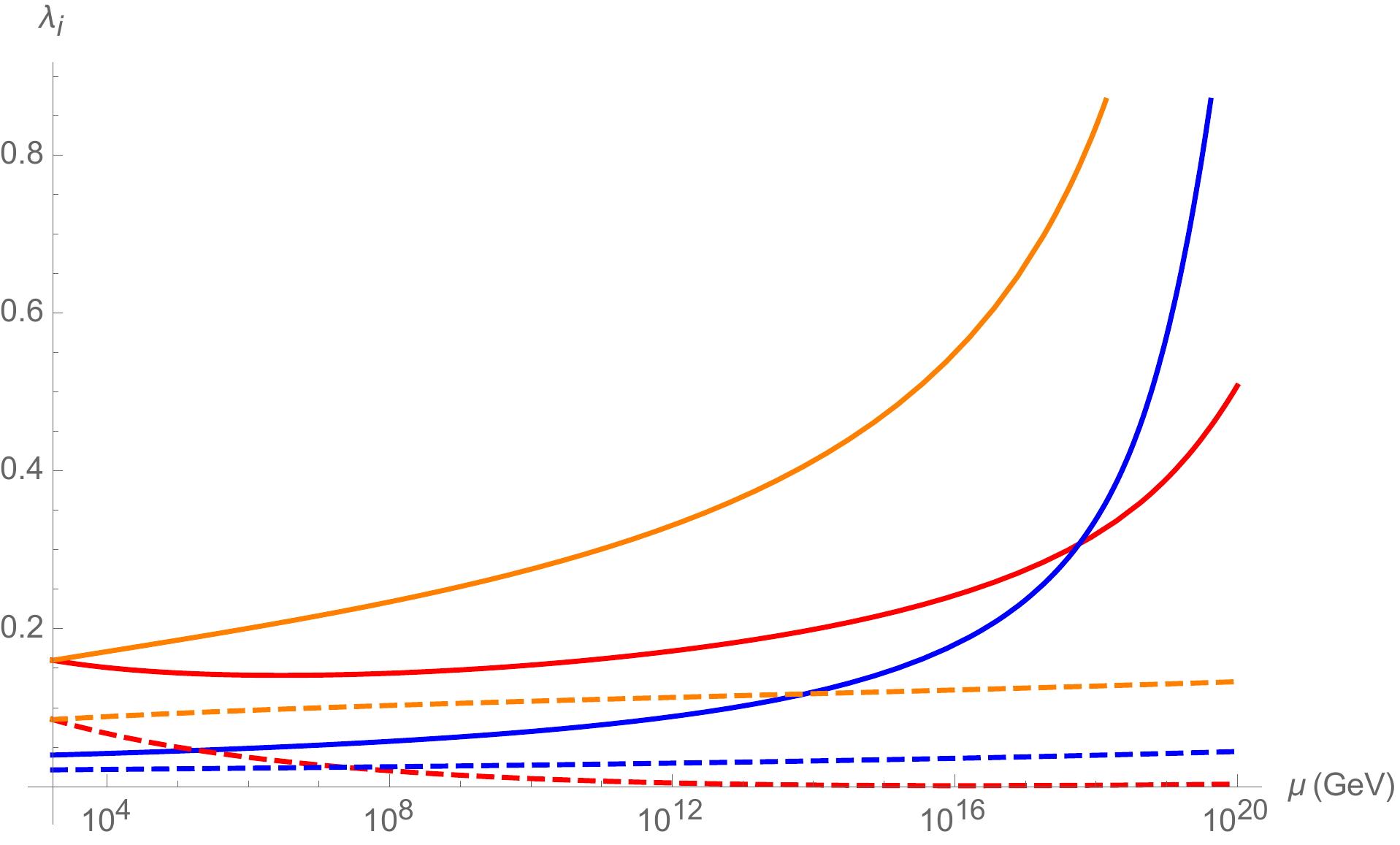}\\[3mm]
	\includegraphics[scale=0.6]{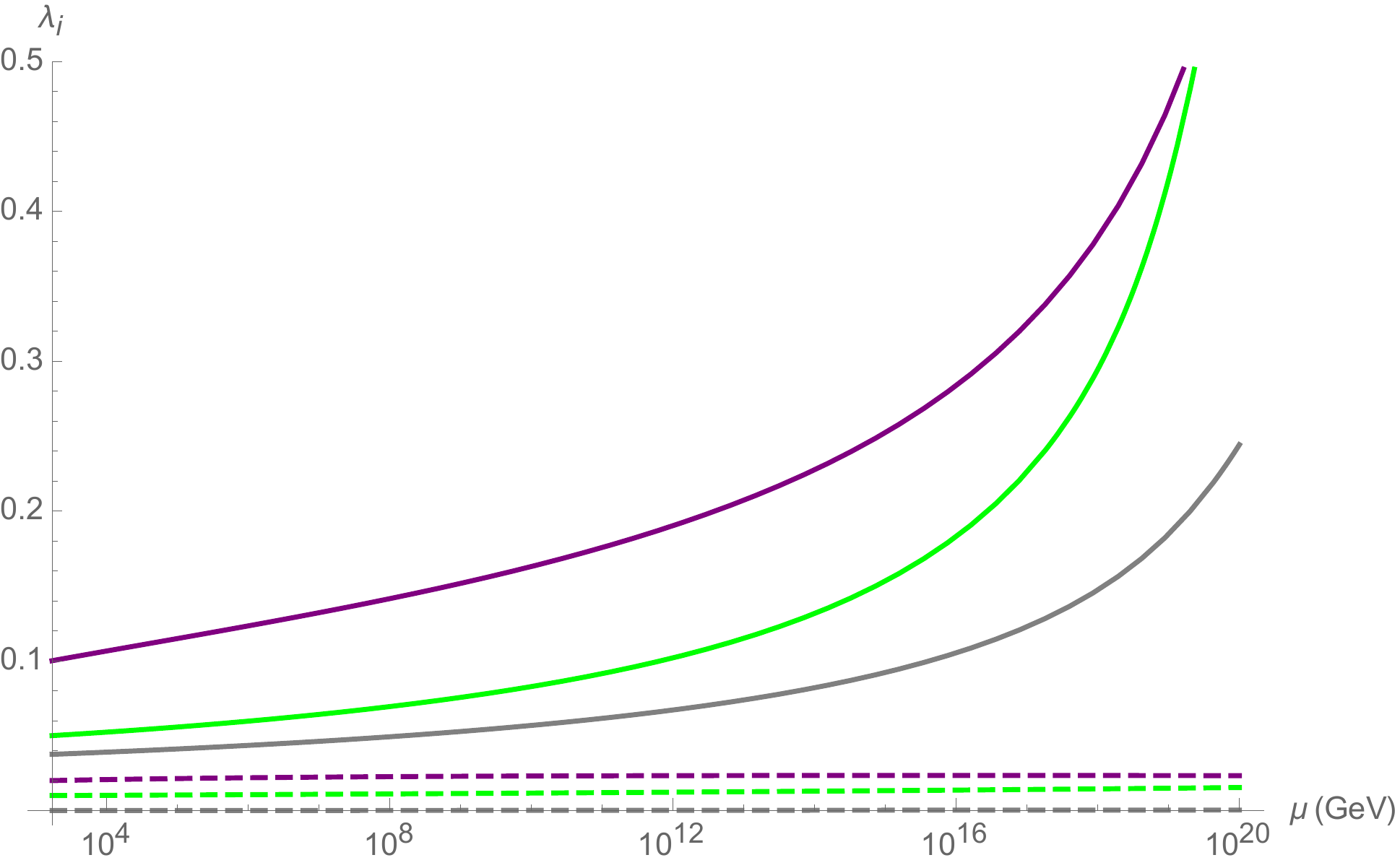}
	\caption{\textbf{Upper panel:} Shows the running of $\lambda_1$ (red), $\lambda_2$ (blue) and $\lambda_3$ (orange) above 1.7 $ \TeV$ where the solid lines correspond to the upper bound on the combined $\lambda$ and the dashed lines are the lover bound. \textbf{Lower panel:} Shows the running of $\lambda_4$ (purple), $\lambda_5$ (green) and $\lambda_\Omega$ (grey) above 1.7 $\TeV$ where the solid lines are the upper bound on the combined $\lambda$ and the dashed lines are the lower bound.}
	\label{fig:Omega}
\end{figure}

The running of the couplings are shown in figure \ref{fig:Omega}. The solid lines on the figure  are from the upper bounds on the couplings and the dashed lines correspond to the lower bounds.

In this case we can calculate the selfcouplings of the Higgs as we did in equation \eqref{eq:sefcoupling} in the previous section. However we find that they are independent of the extra scalar $\Omega$. In the allowed region in the parameter space we find the ratio of the trilinear couplings of the Higgs over the SM value:
\begin{eqnarray}
	1.32 \cdot 10^{-3}< \frac{\lambda_{h_1h_1h_1}}{\lambda_{hhh}^{SM}}<2.43\cdot 10^{-3} \qquad\text{and}\qquad
	18.2 < \frac{\lambda_{h_2h_2h_2}}{\lambda_{hhh}^{SM}}<24.9.
\end{eqnarray}

\section{Conclusion}
\label{sec:conclusion}
We have examined the running of the couplings for the model with a global
symmetry breaking pattern $\SO(5) \to \SO(4)$, which is a minimal
extension of the Standard Model, where the Higgs is an elementary pseudo
Goldstone boson. We have considered the realization of the model in terms
of elementary scalar fields, which is an appealing possibility: the model can
be analysed with controllable perturbative calculations, and the model can in
principle remain valid all the way up to the Planck scale analogously to what
has been proposed to be the case for SM itself. However, when coupled to the
electroweak currents, the model does not provide for a correct symmetry
breaking pattern and vacuum properties \cite{Alanne:2016mmn}.  This issue can be
solved in two different ways. We have separately analysed both of these possiblities to orient the vacuum in the desired and controllable way: First, by adding an explicit symmetry breaking term and second, adding instead an extra singlet scalar field.

To quantify the effects of the running of the couplings, we evaluated the beta
functions of the three couplings in the pure $\SO(5) \to \SO(4)$ model.
The pure doublet coupling depends strongly on the gauge and Yukawa couplings whereas the pure scalar coupling only depends indirectly on these. Adding an explicit breaking term does not change the beta functions and thereby the physical properties of the model.

On the other hand when adding a new scalar, three new couplings emerge and
the three beta functions of the pure $\SO(5)\to\SO(4)$ model are slightly
modified because of the interactions with the new singlet scalar. The new
scalar self-coupling is similar to the original
one as is also the case for the mixing of the new scalar with the doublet.
The coupling between the new and the old scalar is slightly different from the
scalar self-couplings but still it does not depend
directly on the gauge and Yukawa couplings.

Our main result is that the mass intervals of the new heavy scalars
are quite restricted by the overall constraints on the model. 
Below the renormalization scale we required the theory to reproduce the correct mass of the Higgs
and above the renormalization scale we required the couplings to run in such a
way, that the potential remains stable and the couplings remain perturbative all
the way to the Planck scale. In both cases we  analysed in this work we  observed that the running of the
couplings restricts the mass of
the $\sigma$-particle to lie in a narrow interval, $M_\sigma\approx 3$ TeV with the reference renormalization scale to be around $\mu_0\approx 2$ TeV.

More detailed numbers are as follows: When we add an explicit breaking term, the mass lies in the interval $2.93 \TeV \leq M_\sigma \leq 3.10 \TeV$ and the related
renormalization scale is $\mu_0 \approx 1.7\, \TeV$. When we add an extra scalar the
interval of the mass is $2.81 \TeV \leq M_\sigma \leq 2.82 \TeV$ and the
renormlization scale is $\mu_0 \approx 2\, \TeV$. The mass intervals do not overlap
but are close to each other.

We have also compared the trilinear couplings for the Higgs particles and the
heavier scalar state in both cases with
the trilinear coupling of the SM. We found that the trilinear coupling of the
light Higgs is three orders of magnitude smaller than the SM one  and that the
trilinear coupling of the heavy Higgs is one order of magnitude larger in both
cases.

Based on these results, we can expect the models where the Higgs arises as an
elementary pNGB to provide an interesting model building framework which can
be viable up to the Planck scale. The difference with respect to the SM is
the enlarged scalar sector and the dynamical emergence of the electroweak scale
from symmetry breaking at significantly higher energies.

\section*{Acknowledgements}
We thank T. Alanne  for valuable discussions.
H.G., S.G. and F.S.  acknowledges partial support from the Danish National Research Foundation grant DNRF:90. K.T. acknowledges the support from
the Academy of Finland, grant number 267842 and 310130.

\appendix

\section{Broken generators}
\label{app:X5}
First we identify $\SU_L(2)\times \SU_R(2)$ subgroup of $\SO(5)$ and fix the left and right generators as
\begin{equation}
	\left(T_{\mathrm{L,R}}\right)_{ij}^a=-\frac{i}{2}\left[\frac{1}{2}\epsilon^{abc}\left(\delta^b_i\delta^c_j-\delta^b_j\delta^c_i\right) \pm\left(\delta_i^a\delta_j^4-\delta_j^a\delta_i^4\right)\right],
\end{equation}
where the generator $T_R^3$ is identified with the generator of the hypercharge.

The broken generators for $\SO(5)\to \SO(4)$ are then
\begin{equation}
	\label{eq:}
	\begin{split}
		X^{1}_{ij}=&-i\left[\sin\theta \left(\delta_i^1\delta_j^3-\delta_i^3\delta_j^1\right)+\cos\theta\left(\delta_i^1\delta_j^5-\delta_i^5\delta_j^1\right)\right],\\
		X^{2}_{ij}=&-i\left[\sin\theta \left(\delta_i^2\delta_j^3-\delta_i^3\delta_j^2\right)+\cos\theta\left(\delta_i^2\delta_j^5-\delta_i^5\delta_j^2\right)\right],\\
		X^{3}_{ij}=&-i\left[-\sin\theta \left(\delta_i^3\delta_j^4-\delta_i^4\delta_j^3\right)+\cos\theta\left(\delta_i^4\delta_j^5-\delta_i^5\delta_j^4\right)\right],\\
		X^{4}_{ij}=&-i\left(\delta_i^3\delta_j^5-\delta_i^5\delta_j^3\right).
	\end{split}
\end{equation}
The generators are normalised such that $\Tr[X^a_{\theta}X^b_{\theta}]=2\delta^{ab}$.

\end{document}